\documentclass[aps,pra,reprint,superscriptaddress,floatfix,hyperref]{revtex4-1}
\usepackage{bm}
\usepackage{amsmath,amssymb,cases}
\usepackage{graphicx}
\usepackage{color}

\definecolor{mygreen}{RGB}{50, 130, 70}

\newcommand{\fref}[1]{Fig.~\ref{#1}}
\newcommand{\eqnref}[1]{Eq.~(\ref{#1})}
\newcommand{\INO}{Istituto Nazionale di Ottica, CNR-INO, 50019 Sesto Fiorentino, Italy}
\newcommand{\LENS}{\mbox{LENS and Dipartimento di Fisica e Astronomia, Universit\`{a} di Firenze, 50019 Sesto Fiorentino, Italy}}

\begin{document}

\title{Effective expression of the Lee-Huang-Yang energy functional for heteronuclear mixtures}
\author{F. Minardi}
\affiliation{\INO}
\affiliation{\LENS}
\affiliation{Dipartimento di Fisica e Astronomia, Universit\`{a} di Bologna, 40127 Bologna, Italy}

\author{F. Ancilotto}
\affiliation{\mbox{Dipartimento di Fisica e Astronomia `Galileo Galilei' and CNISM, Universit\`{a} di Padova, 35131 Padova, Italy}}
\affiliation{CNR-IOM Democritos, 265–34136 Trieste, Italy}

\author{A. Burchianti}
\affiliation{\INO}
\affiliation{\LENS}

\author{C. D'Errico}
\affiliation{\INO}
\affiliation{\LENS}

\author{C. Fort}
\affiliation{\INO}
\affiliation{\LENS}

\author{M. Modugno}
\affiliation{\mbox{Depto. de Fis\'ica Te\'orica e Hist. de la Ciencia, Universidad del Pais Vasco UPV/EHU, 48080 Bilbao, Spain}}
\affiliation{IKERBASQUE, Basque Foundation for Science, 48013 Bilbao, Spain}

\date{\today}

\begin{abstract}
We consider a homogeneous heteronuclear Bose mixture with contact interactions at the mean-field collapse, i.e. with interspecies attraction equal to the mean geometrical intraspecies repulsion.
We show that the Lee-Huang-Yang (LHY) energy functional
is accurately approximated by an expression that has the same functional form as in the homonuclear case. The approximated energy functional is characterized by two exponents, which can be treated as fitting parameters. We demonstrate that the values of these parameters which preserve the invariance under permutation of the two atomic species are exactly those of the homonuclear case. Deviations from the exact
expression of LHY energy functional are discussed quantitatively and a specific application is described.
\end{abstract}

\maketitle

\section{Introduction}
Bose-Einstein condensates (BEC) with ultracold gases are so dilute that interatomic interactions are generally weak, making the consolidated framework of mean-field (MF) theory
a reliable and accurate approach in many cases. However, even with ultracold gases, there
are instances where quantum fluctuations become relevant, thus producing sizeable deviations from the MF theory.
Going beyond the MF analysis, N. Bogoliubov derived the many-body ground state of a weakly interacting gas of bosons \cite{bogolubov1947}, assuming that most particles occupy the same quantum state, i.e. belong to the condensed fraction, except for a small fraction (``quantum depletion''). The leading correction to the MF energy for identical bosons with hard-sphere interactions was later obtained by Lee-Huang-Yang (LHY) \cite{lee1957} in the homogeneous case. These results were then extended to the case of two-component bosons by D. Larsen  \cite{larsen1963}.

A well-known example where quantum fluctuations are important in ultracold atomic gases is provided by optical lattices that suppress the single-particle kinetic energy and increase the interatomic interactions \cite{Jaksch1998,greiner2002,xu_observation_2006}.
Another case is when the interatomic interactions are enhanced by increasing the scattering length in proximity of a Feshbach resonance. This route was firstly followed in fermionic condensates, where accurate measurements of the frequency shifts in collective oscillations \cite{Altmeyer2007} and of density profiles \cite{Shin2008, Navon2010} showed deviations from the MF predictions. Similar effects
were later observed in strongly interacting bosonic gases in the excitation spectrum of $^{85}$Rb \cite{Papp2008} and in the equation of state of $^7$Li \cite{Navon2011}.
The momentum distribution of the quantum (and thermal) depletion has been
measured in a BEC of metastable He \cite{chang2016} and the total depletion density in a BEC of $^{39}$K \cite{lopes2017}.

Recently, D. Petrov proposed a new setting where quantum fluctuations could have striking effects, i.e. a BEC binary mixture \cite{petrov2015}:
when both components have repulsive intraspecies interactions but the interspecies interactions are attractive and sufficiently large, the MF energy may vanish or even become negative.
In this case, the LHY energy correction dictates the behavior of the system,
allowing for the existence of ``quantum droplets'', i.e. stable states self-bound by interatomic attractive interactions.
Such states
exhibit the properties of 
a liquid even if their density is
several orders of magnitude lower than in ordinary liquids \cite{petrov2015, cappellaro2017, cabrera2018, cheiney2018, semeghini2018, ferioli2019, derrico2019}. Similarly, LHY corrections also stabilize droplets with attractive dipolar forces \cite{ferrier-barbut2016, chomaz2016}.

To describe this novel system, the MF Gross-Pitaevskii equation (GPE) has been modified to incorporate the extra energy due to quantum fluctuations. 
For mixtures, the two-component LHY energy calculated in the homogeneous case is added to the MF energy functional to obtain a generalized GPE (g-GPE), that can be also used in the inhomogeneous case within the local density approximation \cite{cabrera2018, cheiney2018, semeghini2018,ancilotto2018,derrico2019}.
While for an homonuclear mixture of equal mass components the LHY energy functional is analytically known in a closed and simple form \cite{petrov2015}, to date 
no analytic expression is known for a heteronuclear mixture and the LHY term must be calculated numerically \cite{ancilotto2018}.
This implies that the LHY energy must be tabulated for a large number of values of the ratio of two components densities $n_2/n_1$, prior to the numerical integration of the g-GPE.

In this paper, guided by symmetry considerations and asymptotic behavior, we derive a simple and appealing functional form for the heteronuclear mixtures at the MF collapse, as detailed below. We verify that this function
accurately approximates the numerical integral and, as such, can be efficiently used in the numerical solution of the g-GPE and to extend analytical results existing for homonuclear mixtures to the heteronuclear case.

In Section II, after summarizing known results to introduce 
the notation, we review the known analytical LHY energy in the special case of a largely imbalanced mixture, where the lighter species is the majority density component. In other words, this corresponds to the case of heavy impurities
\cite{larsen1963}. In Section III we propose an approximated formula for the LHY energy, which is the main result of this work, and we show its limits of applicability by comparing it with the exact numerical results of the LHY integral. In Secton IV we justify our proposed approximation by means of a Taylor expansion of the LHY integral around the homonuclear case.
In Section V we show the usefulness of an analytic formula with a concrete example: we apply our analytic formula to the case of heteronuclear mixtures
in the so-called LHY ``quantum fluids'' regime, i.e. when the MF interaction is null and the
mixture is governed by the LHY term alone.

\section{LHY energy functional}

The LHY energy of a Bose mixture can be written as
\begin{align}
E_{{\rm LHY}} &= \int d\bm{r} {\cal E}_{\rm LHY}(n_1(\bm{r}),n_2(\bm{r}))\,,
\label{eq:energyfun}
\end{align}
where ${\cal E}_{\rm LHY}$ is the energy density given by \cite{petrov2015,ancilotto2018}
\begin{align}
{\cal E} _{\rm LHY} &= \frac{8}{15 \pi^2} \left(\frac{m_1}{\hbar^2}\right)^{3/2}
\!\!\!\!\!\!(g_{11}n_1)^{5/2}
f\left(\frac{m_2}{m_1},\frac{g_{12}^2}{g_{11}g_{22}},\frac{g_{22}\,n_2}{g_{11}\,n_1}\right)\nonumber
\\
&= \kappa m_{1}^{3/2}(g_{11}n_1)^{5/2}f(z,u,x),
\label{eq:elhy}
\end{align}
with $n_i(\bm{r})$
($i=1,2$) being the density of each component, 
and 
$\kappa\equiv8/(15 \pi^2 \hbar^3)$, $z\equiv m_2/m_1$, $u\equiv g_{12}^2/(g_{11}g_{22})$, $x\equiv g_{22} n_2/(g_{11} n_1)$. The interaction strengths are proportional to the scattering lengths $a_{11}, a_{22}, a_{12}$: $g_{ii}=4\pi\hbar^2 a_{ii}/m_i, g_{12}=2\pi\hbar^2 a_{12}/m_{12}$, with $m_{12}$
the reduced mass. We note that the LHY energy density is independent of the sign of $g_{12}$, but the formation of quantum droplets occurs only for negative $g_{12}$. Therefore, as done in Ref. \cite{petrov2015}, here we consider the case of a mixture with interspecies attractive interaction, namely $g_{12}<0$, with $g_{11}>0$ and $g_{22}>0$.

For the homonuclear case, $z=1$, the function is given in Ref. \cite{petrov2015}:
\begin{equation}
\label{eq:f_homo}
    f(1,u,x) = \sum_{\pm}\left( 1+x\pm\sqrt{(1-x)^2+4ux} \right)^{5/2}/(4\sqrt{2})
\end{equation}
so that $f(1,1,x) = (1+x)^{5/2}$.
For the heteronuclear case, $z\neq 1$, a simple expression similar to the one above is not available and approximate or numerical methods are needed.

The expression of $f(z,u,x)$ for heteronuclear mixtures is obtained from Ref. \cite{petrov2015, ancilotto2018}:
\begin{equation}
f(z,u,x) = \frac{15}{32} \int _0^\infty  k^2 {\cal F}(k,z,u,x)  \, dk \,,
\label{eq:integral_f_u}
\end{equation}
\begin{widetext}
\begin{eqnarray}
\label{eq:long_integral}
{\cal F}(k,z,u,x) &=&
\left\{ \frac{1}{2}\left[k^2 \left(1+
\frac{x}{z}\right)+\frac{1}{4} k^4 \left(1+\frac{1}{z^2}\right) \right]
+ \left[ \frac{1}{4} \left[\left(k^2+\frac{1}{4}k^4\right)-
\left(\frac{x}{z}k^2+\frac{1}{4z^2}\,k^4\right)\right]^2 + u  \frac{x}{z}\,k^4 \right]^{1/2}\right\}^{1/2}
\\
\nonumber
&+& \left\{ \frac{1}{2}\left[k^2 \left(1+\frac{x}{z}\right)+\frac{1}{4} k^4 \left(1+\frac{1}{z^2}\right) \right]
- \left[ \frac{1}{4} \left[\left(k^2+\frac{1}{4}k^4\right)-
\left(\frac{x}{z}k^2+\frac{1}{4z^2}\,k^4\right)\right]^2 +
u\frac{x}{z}\,k^4 \right]^{1/2}\right\}^{1/2}
\\
 \nonumber
&-&\frac{1+z}{2z}\, k^2 -(1+x)+\,\frac{1}{k^2}  \left[1+x^2 z+4 u x\frac{z}{1+z} \right] \,.
\end{eqnarray}
\end{widetext}

Since the dimensionless function $f(z,u,x)$ is weakly dependent on 
$u$, as shown in Ref. \cite{petrov2015}, the latter is usually replaced by its value at MF collapse, where $g_{12}+\sqrt{g_{11}g_{22}}=0$, i.e. $u=1$:
\begin{align}
\label{eq:elhy2}
{\cal E} _{\rm LHY} &\simeq  \kappa m_{1}^{3/2}(g_{11}n_1)^{5/2}f(z,1,x).
\end{align}

The expression of $f(z,1,x)$ coincides with that given in Ref. \cite{ancilotto2018}, where it is remarked that, since
the integral converges due to the cancellation of divergent terms, its evaluations requires some care.

\subsection{Limit of heavy impurities}
Lacking a closed form like \eqnref{eq:f_homo}, an analytic expression of the LHY energy is given in Ref.~\cite{larsen1963} for the special case of a repulsive mixture with $a_{11}=a_{12}=a_{22}>0$, where the density of the heavy component is much lower than that of the light component, i.e. $z>1$ and $x \ll 1$:
\begin{align}
\label{eq:larsen}
{\cal E} _{\rm LHY} &= \frac{8}{15 \pi^2} \left(\frac{m_1}{\hbar^2}\right)^{3/2}
\!\!\!\!\!\!(g_{11}n_1)^{5/2}
f_L(z,x)\\
 f_L(z, x)
  &= 1+ x \frac{15}{16} \frac{z+1}{z-1}  \left[  \frac{z^2}{\sqrt{z^2-1}} \arctan \sqrt{z^2-1} -1\right] \label{eq:larsen1}
\end{align}

Even if the above equation was derived for all positive scattering lengths, we can also apply it to the case of $-a_{12}=a_{11}=a_{22}>0$ since the LHY energy is independent of the sign of $a_{12}$, as shown by \eqnref{eq:elhy}. Note, however, that the condition of all equal scattering lengths corresponds to $u= (1+z)^2 /(4z)$, thus $u=1$ only for $z=1$.

In the general case, however, the numerical integration of the g-GPE requires the knowledge of the LHY term for generic values of $x$, and the above \eqnref{eq:larsen} is obviously insufficient. However, we will use it to confirm the validity of our approximated formula proposed below. For the purpose, it is useful to consider $f_L(z,x)$ around $z=1$, where it can be compared to $f(z,1,x)$:

\begin{equation}
\label{eq:larsenTaylor}
 f_L(z, x) \simeq 1+ \frac52 x + \frac32 x(z-1) + O[z-1]^2.
\end{equation}

\section{Effective Ansatz}
\begin{figure}[t]
\centerline{\includegraphics[width=0.9\columnwidth]{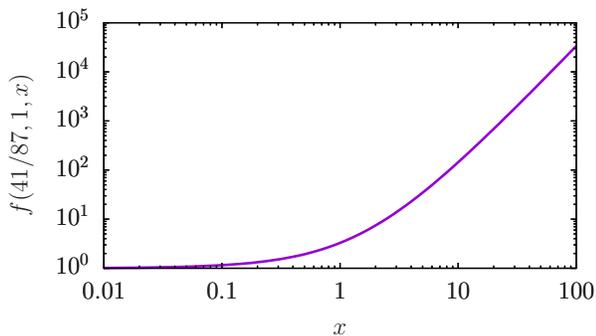}}
\caption{Values of $f(z,1,x)$ as a function of $x$, for $z=41/87$, as obtained from numerical integration of \eqnref{eq:integral_f_u}}
\label{fig1}
\end{figure}

We introduce here the main result of our work, namely an approximate analytic expression for $f(z,1,x)$. The exact function, defined by Eqs. (\ref{eq:integral_f_u}, \ref{eq:long_integral}), can only be evaluated by points, through numerically integration of the above equations. As an example, in \fref{fig1} we plot $f(41/87,1,x)$, corresponding to the case of a  $^{41}$K-$^{87}$Rb mixture.

When seeking for an approximate 
expression for $f(z,1,x)$,
it is useful to
recall that in the case of a single component the LHY correction takes the form \cite{lee1957,dalfovo1999}
\begin{equation}
{\cal E}_{{\rm LHY}}=\kappa m^{3/2}\left(g n\right)^{5/2}.
\end{equation}
Thus, since for $x \to 0$ the LHY energy has to converge to the value of the single species 1,
in this limit $f(z,1,x) \to 1$.
In the opposite limit, i.e. for $x\gg1$, the LHY energy converges to the value of the single species 2, ${\cal E}_{{\rm LHY}}\to \kappa m_2^{3/2} (g_{22} n_2)^{5/2}$, implying $f(z,1,x) \to z^{3/2} x^{5/2}$.

Having established the above asymptotic behavior, we also recall that ${\cal E}_{{\rm LHY}}$ must be invariant under permutation of the two species. It is therefore convenient to rewrite Eq. (\ref{eq:elhy}) in a more symmetric form with respect to the species index:
\begin{equation}
{\cal E}_{{\rm LHY}}=\kappa (m_1 m_2)^{3/4} (g_{11} n_1 g_{22}n_2)^{5/4} z^{-3/4}\!\! x^{-5/4}  f(z, 1, x).
\label{eq:eLHYxz}
\end{equation}
Now it is easy to see that invariance under permutation of indices $i=1,2$ implies
\begin{equation}
f(z,u,x) = z^{3/2} x^{5/2} f\left(\frac{1}{z},u,\frac{1}{x}\right).
\label{eq:swap}
\end{equation}

\begin{figure}
\centerline{\includegraphics[width=0.9\columnwidth]{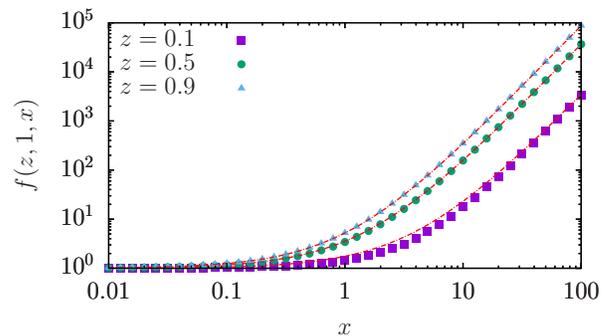}}
\caption{Numerical values of integrals (points) compared to the approximate function
$(1+z^{3/5}x)^{5/2}$ (red dotted-dashed lines), for $z=0.9, 0.5, 0.1$. We restrict to $z\le1$ without loss of generality
.}
\label{fig:simpleFit}
\end{figure}

The above properties suggest that one could approximate
$f(z,1,x)$  with the following function
\begin{equation}
F_{q}(x, z) = \left[1 + \left(z^{3/2} x^{5/2}\right)^{1/q}\right]^q ,
\label{eq:fitfun}
\end{equation}
which has a functional form that resembles that of the homonuclear case \eqnref{eq:f_homo}, except for the presence of the factor $z^{3/2}$, accounting for the mass ratio,
and of the exponent $q (1/q)$, that is here considered as a fitting parameter. In the specific case
shown of Fig. \ref{fig1}, a best fit to $f(z,1,x)$ returns $q=2.4612(2)$
(with the relative residuals being below $2\%$ for all values of $0.01<x<100$). In general the $q$ parameter returned by the fit depends on $z$,
but for a broad range of $z$ values
 it is very close to the value $q= 5/2$, which characterizes the function $f$ in the homonuclear mixture, \eqnref{eq:f_homo} \cite{fitParameters}.
%
%
Thus, we find convenient to approximate
$f(z,1,x)$ with the following, simple generalization of the expression for
the homonuclear case:
\begin{equation}
\label{eq:simple_f}
f\left(z,1,x\right)\simeq F_{5/2}(z,x) \equiv (1 + z^{3/5} x)^{5/2}.
\end{equation}
As shown in Fig. \ref{fig:simpleFit}, this form, in spite of its simplicity,
provides a very accurate representation of the exact (numerical) values of $f(z,1,x)$, in a wide range of mass ratios.
In addition, as requested, \eqnref{eq:simple_f} is consistent with above \eqnref{eq:larsenTaylor} when we consider $x, |z-1| \ll 1$.
\begin{figure}[!t]
\centerline{\includegraphics[width=0.9\columnwidth]{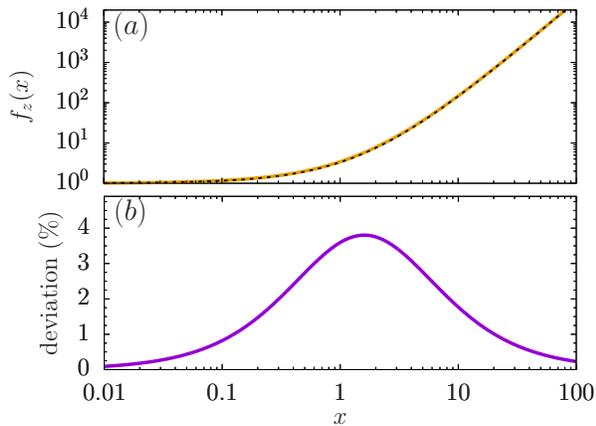}}
\caption{Top: plot of $F_{5/2}(z,x)$ (solid line) and $f^{(3)}(z,x)$ (dashed line), as defined in text.
Bottom: relative deviation $|f^{(3)}(z,x)-F_{5/2}(z,x)|/f^{(3)}(z,x)$ (percentage). Here $z=41/87$.}
\label{fig:superserie}
\end{figure}

Finally, by combining \eqnref{eq:elhy2} and \eqnref{eq:simple_f}, we arrive at the following \textit{effective expression} for the LHY energy functional
\begin{equation}
\label{eq:elhy3}
{\cal E} _{\rm LHY} \simeq  \frac{8m_{1}^{3/2}(g_{11}n_1)^{5/2}}{15 \pi^2 \hbar^3}
\left[1 + \left(\frac{m_2}{m_1}\right)^{3/5}
\frac{g_{22} n_2}{g_{11} n_1}\right]^{5/2}\!\!\!\!\!\!\!\!,
\end{equation}
which constitutes one of the main results of this work.

\section{Perturbative expansion}

In order to quantify the deviation of the exact
$f(z, 1, x)$ from $F_{5/2}(z,x)$, 
we note that the latter is only function of $\xi\equiv z^{3/5} x$ (and not of $x$ and $z$ separately). Then we define the auxiliary function
\begin{equation}
g(z,\xi)\equiv f(z, 1, z^{-3/5}\xi);
\end{equation}
if $f(z,1,x)$ was a function of $\xi$ alone, then $g(z,\xi)$ would be independent on $z$. Although this is not 
the case in general, close to $z=1$,  $g(z,\xi)$ is almost independent of z.
To show this, we write the Taylor expansion:
\begin{equation}
g(z,\xi)=\sum_{n=0}^{+\infty}G_{n}(\xi)(z-1)^n,
\end{equation}
and compute analytically the coefficients $G_{n}$
from the integral \eqnref{eq:long_integral}

\begin{align}
G_{0}(\xi)&=(1 + \xi)^{5/2}
\\
G_{1}(\xi)&=0
\\
G_{2}(\xi)&=-\frac{12}{35}\xi\sqrt{1+\xi}
\\
G_{3}(\xi)&= \frac{4 \xi  (47 \xi +43)}{525 \sqrt{\xi +1}}.
\end{align}

The vanishing first derivative, $G_1(\xi)=0$, confirms that $g(z,\xi)$ depends weakly on $z$ around $z=1$. Therefore,
\begin{align}
f(z, 1, x) &= (1 + z^{3/5} x)^{5/2} -\frac{12}{35}z^{3/5} x\sqrt{1+z^{3/5} x}(z-1)^{2}
\nonumber\\
&+ \frac{4 z^{3/5} x  (47 z^{3/5} x +43)}{525 \sqrt{z^{3/5} x +1}}(z-1)^{3}
+ O\left((1-z)^{4}\right)
\nonumber\\
&\equiv f^{(3)}(z,x)+O\left((z-1)^{4}\right)
\label{eq:superserie}
\end{align}

The above equation shows that our ansatz function $F_{5/2}(z,x)$ is the zeroth-order term of the Taylor expansion around $z=1$.
A comparison between $F_{5/2}(z,x)$ and the third-order truncated Taylor series, $f^{(3)}(z,x)$, is shown in Fig. \ref{fig:superserie} for $z=41/87$. The lower panel shows that the first term, $F_{5/2}(z,x)$, approximates $f(z, 1, x)$ better for small and large values of $x$, as expected since it 
is determined to match the asymptotic behaviours. Anyway, the maximum relative deviation is below 4\%, proving that $F_{5/2}(z,x)$ suffices for a reliable estimation of the LHY integral.
\section{A Lee-Huang-Yang fluid with heteronuclear mixtures}

Recently the so called ``Lee-Huang-Yang
fluid" has been proposed, consisting in a dilute quantum mixture where mutual
interactions and atom numbers are tuned such that the
MF interactions cancel out, and the resulting system is
governed only by quantum fluctuations \cite{jorgensen2018, beyondLHY}.
This system has been studied in the case of a homonuclear mixture
($m_1=m_2$) in Ref. \cite{jorgensen2018}, where explicit expressions
for important quantities such as, e.g.,
the healing length 
and the frequency of monopole oscillation frequency, are obtained.

To prove the usefulness of our analytical, albeit
approximate, form for the LHY energy discussed in
Section III, we provide here
a generalization of results derived in Ref. \cite{jorgensen2018}
to the experimentally relevant case of heteronuclear mixtures, by explicitly using our ansatz
formula \eqnref{eq:simple_f}.

The total energy of a Bose-Bose binary mixture, 
with densities
$n_1,n_2$ and interacting
via contact potentials, is given by
\begin{widetext}
\begin{equation}
E=\int d{\bf r}\left\{ \sum _{i=1,2} \left[ \frac{\hbar^2}{2m_i}|\nabla \psi _i|^2
+ V_i |\psi _i|^2 + \frac12g_{ii}|\psi _i|^4   \right]
 + g_{12}|\psi _1|^2|\psi_2|^2  +  \cal{E}_{\rm LHY} \right\}
\label{eq:elhy1}
\end{equation}
\end{widetext}
where $\psi_i$ and $V_i$ are the wavefunction and the external potential of the $i-$th component ($i=1,2$), respectively, and $\cal{E}_{\rm LHY}$ is
the LHY energy density contribution \eqnref{eq:elhy}.
We assume in the following a symmetric harmonic trap,
$V_1({\bf r})=V_2({\bf r})=m_1\omega _1^2r^2/2\equiv V_{ho}$.
The associated harmonic length is $a_{ho}=\sqrt{\hbar/m_1 \omega_1}$.

The atom numbers and interactions can be tuned to cancel
completely the mean-field terms in the above energy functional.
In fact, by choosing $g_{12}=-\sqrt{g_{11}g_{22}}$
(namely, the MF collapse condition), and atom numbers such that
$N_2/N_1=\sqrt{g_{11}/g_{22}}$,
only the LHY energy term remains in \eqnref{eq:elhy1},
where the two component wavefunctions satisfy the condition $\psi _2=\psi _1 (g_{11}/g_{22})^{1/4}$.

By defining $|\psi |^2=|\psi _1|^2+|\psi _2|^2$
and $\alpha = \sqrt{g_{11}/g_{22}}=z^{1/2}\sqrt{a_{11}/a_{22}}$,
one has $|\psi _1|^2=(1+\alpha)^{-1}|\psi |^2$
and $|\psi _2|^2=\alpha (1+\alpha)^{-1}|\psi |^2$. Thus the
total energy functional \eqnref{eq:elhy1} reduces to:
\begin{widetext}
\begin{align}
E&= \int  d{\bf r} \left[\frac{\hbar^2}{2}|\nabla \psi |^2 \left(\frac{1}{m_1(1+\alpha)}
+ \frac{\alpha}{m_2(1+\alpha)}\right)+
V_{ho} |\psi |^2+ \cal{E}_{\rm LHY}
\right]
=\int  d{\bf r} \left[\frac{\hbar^2}{2m^\ast }|\nabla \psi |^2+
V_{ho} |\psi |^2 + \cal{E}_{\rm LHY}
\right]
\label{eq:elhy_red}
\end{align}
\end{widetext}
where
\begin{align}
m^* = m_1 \,\left[\frac{1+\sqrt{za_{11}/a_{22}}} {1+\sqrt{a_{11}/za_{22}}}\right].
\label{eq:emass}
\end{align}
Then, by approximating $\cal{E}_{\rm LHY}$
by the effective expression in \eqnref{eq:elhy3}, we can write
\begin{align}
\cal{E}_{\rm LHY}&= \frac{8}{15 \pi^2} \left(\frac{m_1}{\hbar^2}\right)^{3/2}\left(\frac{g_{11}n}{1+\alpha}\right)^{5/2}
\!\!\left(1 + z^{3/5} \alpha^{-1/2}\right)^{5/2}
\nonumber \\
&=C^*_{{\rm LHY}} |\psi|^{5}
\label{eq:elhy4}
\end{align}
with
\begin{equation}
C^*_{{\rm LHY}} \equiv \frac{256\sqrt\pi}{15}\frac{\hbar^2}{m_1} \left[a_{11}
\frac{ 1+z^{1/10} \sqrt{a_{22}/a_{11}}}
{  1+ \sqrt{za_{11}/a_{22}}}\right]^{5/2}.
\end{equation}
The expression in \eqnref{eq:elhy4} makes the above \eqnref{eq:elhy_red}
formally identical to the
total energy functional for the homonuclear LHY fluid of
Ref.~\cite{jorgensen2018}, the only difference being that the coefficient
$256\sqrt{\pi}\hbar ^2|a_{12}|^{5/2}/(15m)\equiv C_{{\rm LHY}}$
is replaced by $C^*_{{\rm LHY}}$, $m$ by $m^*$,
and $\omega_1$ by $\omega^*\equiv\sqrt{m_1/m^*}\,\omega _1$.

With the above replacements the results of Ref.~\cite{jorgensen2018} can be easily generalized to the case of different masses, $z\neq 1$. For instance,
we readily obtain the healing length in the homogeneous case, $V_{ho}(\mathbf{r})=0$,
\begin{align}
    \xi_{{\rm LHY}}^{-2} &= \frac{256 \sqrt{\pi} n^{3/2} a_{11}^{5/2}}{3}
    \left[ \frac {1+\sqrt{z a_{11}/a_{22}}}{1+\sqrt{a_{11}/z a_{22}}}
    \right]
    \times \nonumber \\
    & \quad \quad
    \left[
    \frac{  1+z^{1/10}\sqrt{a_{22}/a_{11}}}
    {  1+ \sqrt{za_{11}/a_{22}}  }\right]^{5/2},
\end{align}
and the frequency $\omega$ of monopole collective oscillation in the weakly interacting limit, i.e. $N^{3/2}|a_{12}/a_{ho}|^{5/2}\ll 1$ (where $N=N_1+N_2$ is the total atom number):

\begin{widetext}
\begin{equation}
    \omega / \omega ^* =2 +  \frac{64\sqrt2 N^{3/2}}{5\sqrt5\,\pi^{7/4}}
    \left(\frac{a_{11}}{a_{ho}}\right)^{5/2}
    \left[
    \frac{ 1+\sqrt{za_{11}/a_{22}} } {1+\sqrt{a_{11}/za_{22}} }
    \right]^{13/8}
    \left[
    \frac{  1+z^{1/10}\sqrt{a_{22}/a_{11}}}
    {  1+ \sqrt{za_{11}/a_{22}}  }
    \right]^{5/2}.
\end{equation}
\end{widetext}
As expected, in the case of equal masses $z=1$ the above expressions reduce to the corresponding expressions (10) and (15) of Ref.~\cite{jorgensen2018}.

In order to estimate the effect of mass imbalances on the properties of the LHY fluid, we evaluate the deviation of the monopole oscillation frequency from the case of ideal gases, defined as $(\omega/\omega^{*} - 2)/N^{3/2}\equiv B$, for various heteronuclear Bose mixtures which are currently realized in experiments of tunable dual-species condensates, as summarized in Table~\ref{tab:BBexps}. Experimentally, the frequencies of collective modes oscillations are measured with high accuracy and have been used throughout as a powerful tool to study many-body dynamics. With typical atom numbers of $N\sim10^4$, the deviations $\omega/\omega^* -2$ should be readily measurable.

In the Thomas-Fermi regime, i.e. the opposite limit $N^{3/2}|a_{12}/a_{ho}|^{5/2}\gg 1$, we obtain the total peak density from Eq. (9) of Ref.~\cite{jorgensen2018}:
\begin{align}
    n_0 & = \frac{A^{4/3} 3^{2/3}}{32(2\pi)^{1/3} } \left(\frac{N}{a_{ho}^{*6}}\right)^{4/13} \left(
    \frac{15 m^* C^*_{LHY}} {256\sqrt\pi\hbar^2}\right)^{-6/13}\nonumber\\
& \simeq 0.078  \left(\frac{N}{a_{ho}^6}\right)^{4/13}
\left[ a_{11}
\frac{  1+z^{1/10}\sqrt{a_{22}/a_{11}}}
    {  1+ \sqrt{za_{11}/a_{22}}  }
    \right]^{-15/13}
    \end{align}
where $A\simeq 1.815$. This quantity is relevant for the lifetime of the atomic LHY quantum fluid that is limited by 3-body inelastic collisions, whose rate is proportional to $n_0^2$. In Table~\ref{tab:BBexps} we report also the peak densities evaluated for the different Bose mixtures.

\begin{table}[t!]
    \centering
    \begin{tabular}{|c|c|c|c|c|c|c|c|}
    \hline
         \multicolumn{2}{|c|}{Species 1,2} & $a_{11}/a_0$ & $a_{22}/a_0$ & $z$& $10^7 B$& $n_0/N^{4/13}$[cm$^{-3}$] & Ref.\\ \hline
         $^{39}$K& $^{39}$K & 86 & 34 & 1 & 2.4 &$4.0 \cdot 10^{13}$ &\cite{cabrera2018},\cite{semeghini2018} \\
         $^{23}$Na& $^{39}$K & 52 & 13 & 1.7 & 0.32 & $8.9\cdot 10^{13}$ &\cite{schulze2018} \\
         $^{41}$K& $^{87}$Rb & 65 & 100.4 & 2.1 & 8.2 & $3.1\cdot 10^{13}$ & \cite{derrico2019} \\
         $^{39}$K& $^{87}$Rb & 10 & 100.4 & 2.2 & 0.76 & $7.9\cdot 10^{13}$ &\cite{wacker2015}\\
         $^{23}$Na& $^{87}$Rb & 54.5 & 100.4 & 3.8 & 5.0 & $3.0\cdot 10^{13}$ & \cite{wang2016} \\
         \hline
    \end{tabular}
    \caption{Bose mixtures experimentally realized and amenable to collapse at $g_{12}=-\sqrt{g_{11}g_{22}}$. For the purpose of comparison, we assume the same harmonic potential for all species, corresponding to an harmonic frequency of $100$ Hz for $^{87}$Rb, i.e. $\omega_j/(2\pi) = 100 \sqrt{m_{Rb}/m_j}$ Hz for species $j$.}
    \label{tab:BBexps}
\end{table}

\section{Conclusions}
We have shown that
the Lee-Huang-Yang energy functional for a heteronuclear Bose mixture at the MF collapse can be accurately approximated by an expression that has a functional form similar to the simple expression of the homonuclear case, see Eq. (\ref{eq:elhy3}). Two different ansatz functions, with one or two exponents as fitting parameters, have been considered \cite{fitParameters}. Remarkably, we found that the values of these fitting parameters which preserve the invariance under permutation of the two species are exactly  those of the homonuclear case. A quantitative analysis of the deviations from the exact expression of LHY energy functional indicates that the simple expression we propose is effective in a wide range of mass ratios, and may be useful for describing current and future experiments. Our formula greatly simplifies the numerical integration of the generalized GPE's and allows further analytic study, as demonstrated in the case of a Lee-Huang-Yang fluid with heteronuclear mixtures.

\begin{acknowledgments}
We are grateful to M. Prevedelli and L. Salasnich for comments and suggestions, and to L. Marmugi for carefully reading the manuscript.
We acknowledge support by the Spanish Ministry of Science, Innovation and Universities and the European Regional Development Fund FEDER through Grant No. PGC2018-101355-B-I00 (MCIU/AEI/FEDER,UE), the Basque Government through Grant No. IT986-16, and the Fondazione CR Firenze through project ``SUPERACI-Superfluid Atomic Circuits''.
\end{acknowledgments}

\bibliography{droplet}

\end{document}